\documentclass[twocolumn,showpacs,amsmath,amssymb,prl,superscriptaddress]{revtex4}
\usepackage{amssymb}
\usepackage{natbib}
\usepackage{graphicx}

\begin{document}

\title{Phase transition in a static granular system}

\date{\today}

\author{Matthias Schr\"oter}
\email{schroeter@chaos.utexas.edu}
\affiliation{Center for Nonlinear Dynamics and Department of Physics, The University of Texas at Austin, Austin,
         Texas 78712, USA}
\author{Sibylle N\"agle}
\affiliation{Center for Nonlinear Dynamics and Department of Physics, The University of Texas at Austin, Austin,
         Texas 78712, USA}
\author{Charles Radin}
\email{radin@math.utexas.edu}
\affiliation{Mathematics Department University of Texas at Austin, Austin TX 78712}

\author{Harry L. Swinney}
\affiliation{Center for Nonlinear Dynamics and Department of Physics, The University of Texas at Austin, Austin,
         Texas 78712, USA}

\date{\today}

\begin{abstract}
We find that a column of glass beads exhibits a well-defined transition
between two phases that differ in their resistance to shear. Pulses of
fluidization are used to prepare static states with well-defined
particle volume fractions $\phi$ in the range 0.57-0.63.  The
resistance to shear is determined by slowly inserting a rod into the
column of beads.  The transition occurs at $\phi=0.60$ for a
range of speeds of the rod.

\end{abstract}


\maketitle

A static assembly of granules, for instance sand in a rigid container,
responds differently to shear when packed loosely from when packed
tightly \cite{nedderman:92}.  It is natural to enquire whether these two states are
smoothly connected as volume fraction varies, or, as with assemblies
of particles in thermal equilibrium, are such states sharply separated
by one or more phase transitions.  We use recent advances in
controlling the preparation of granular assemblies to show that the
latter holds.

An old magic trick is based on the qualitative difference in
the resistance to shear of loosely packed and tightly packed particles:
When a pot with a narrow neck is loosely filled with grains, a rod is
easily inserted and withdrawn. The rod is then inserted and the grains
are shaken or otherwise agitated to a denser state, whereupon the
whole apparatus can be lifted by the rod and spun about the
performer's head \cite{branson:22, aste:00}.

The existence of distinct phases in granular matter has been widely
discussed, but a sharp distinction between the two phases has remained
elusive \citep{jaeger:96,kadanoff:99,gennes:99}, the distinction being
hampered by the difficulty in preparing a well-defined initial state
\citep{gennes:99}.  The effort to overcome this was advanced
significantly by Nowak {\it et al.}~\cite{nowak:98}, who used a
mechanical tapping protocol to obtain well-defined volume fractions
$\phi$ in the range 0.628 -- 0.658.  Recently Schr\"oter {\it et al.}
\cite{schroeter:05} showed that a protocol based on expanding the
granular medium by pulses of fluid from below could be used to prepare
a column of grains with $\phi$ defined to within 0.1\%. Using this
technique, we prepare granular samples in the range 0.571 $<$ $\phi$
$<$ 0.633.

{\it Experiment.---} To measure the force of resistance to the insertion
of a rod into a granular sample we use an apparatus similar to that in
\cite{stone:04,stone:04b,hill:05}. Those previous studies focused on the
influence of geometrical factors such as the size of particles, rod,
and vessel, and on how the penetration force increased when the rod
approached the bottom boundary. Those experiments were
performed at a single volume fraction, $\phi$ = 0.59 \cite{stone:04,stone:04b}.

Our measurements are performed with a home built granular
penetrometer: a translation stage (driven by a stepper motor with a
step size 2.5 $\mu$m) moves a stainless steel rod (diameter 6.3 mm and
flat head) downwards into a granular sample. The force needed for
penetration is measured with a load cell with a full range of 10 N
(Honeywell, Model 31). The sample consists of soda lime glass beads
from Cataphote with a diameter of 265 $\pm$ 15 $\mu$m and a density of
2.484 $\pm$ 0.002 g/cm$^3$ (measured with a Micromeritics gas
pycnometer AccuPhys 1330). The beads are
contained in a water-fluidized bed where flow pulses of different flow
rates allow us to select a volume fraction $\phi$
for the static sedimented bed \cite{schroeter:05}. (If air rather than water is used to
fluidize a bed, it is difficult to obtain low enough volume fraction
to see the transition \cite{ojha:00}.)  The beads are fluidized inside
a square bore glass tube (39.9 $\times$ 39.9 mm$^2$).  The ratio of
inner tube size to rod diameter is 6.3, larger than the value five
that \cite{stone:04} found to be sufficient so that the influence of
the vessel walls was negligible.  Flow pulses are generated using a
digital gear pump (Barnant Co., model no. 75211). The volume fraction
is determined from measurement of the bed height; one pixel in the
digital images corresponds to a change of only $0.02\%$ in $\phi$.

\begin{figure}[t]
  \begin{center}
    \includegraphics[angle=-90,width=8.8cm]{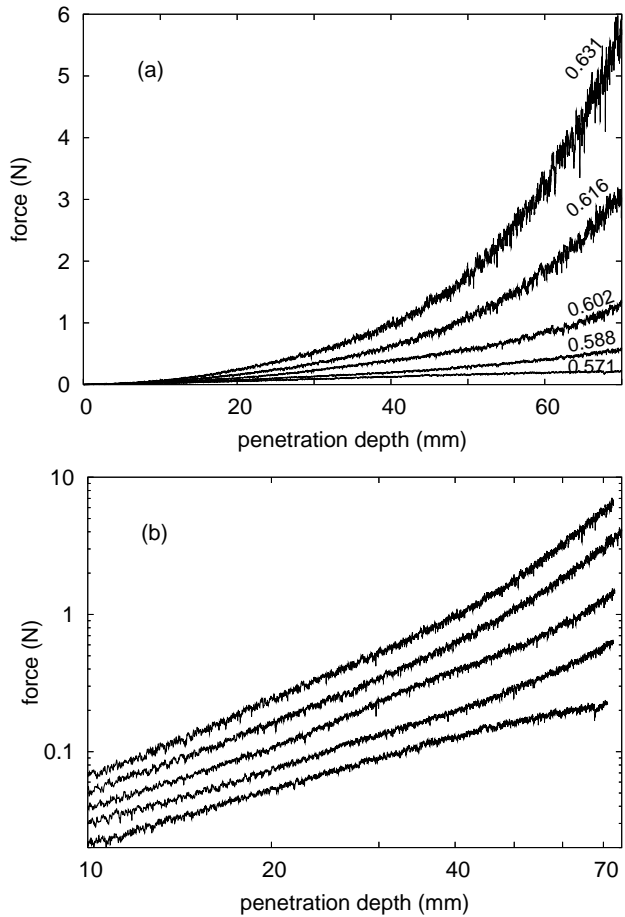}
    \caption{(a) Penetration force as a function of depth for
        different volume fractions. (b) Double logarithmic plot of the
        data in (a).  These curves are measured with a penetration speed
        of 10 mm/s, but the results are the same for a range of
        penetration speeds. The total
        sample height is 110 mm (at $\phi$ = 0.6).  }
    \label{fig:force}
  \end{center}
\end{figure}

{\it Results.---} The force on the penetrating rod increases
monotonically with its depth, as Fig.~\ref{fig:force} (a) illustrates.
Contrary to the observations in \cite{hill:05}, we find that the rate
of growth is not polynomial (see the double logarithmic plot in
Fig.~\ref{fig:force} (b)). Further, at maximum depth the rod is 40 mm
from the bottom of the container, well beyond the distance of 20 mm
where boundary effects have been found to be measurable for conditions
comparable to our experiment \cite{stone:04}.

\begin{figure}[t]
  \begin{center}
    \includegraphics[angle=-90,width=8.8cm]{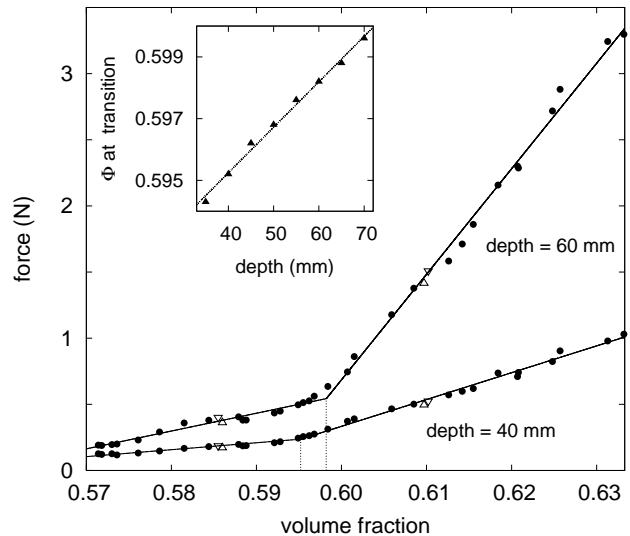}
    \caption{The dependence of the penetration force on volume
fraction changes at $\phi = 0.598$ at a depth of 60 mm.
Data represented by $\bullet$ were
measured at a penetration speed of 10 mm/s, $\triangledown$ at 5
mm/min and $\triangle$ at 20 mm/min.  The transition points are determined 
by the intersection of least-square fits for the volume fractions 
below 0.595 and above 0.6. The inset displays the dependence of the 
transition point on the penetration depth. 
}
    \label{fig:transition}
  \end{center}
\end{figure}

The rate of change of force with volume fraction exhibits a
well-defined transition, which occurs for $\phi$ = 0.598  for a rod depth of 60 mm, as
Fig.~\ref{fig:transition} illustrates.  The $\phi$ values
corresponding to the transition increase slowly with rod depth (cf.
inset of Fig.~\ref{fig:transition}). This dependence of the transition
point on the penetration depth cannot correspond to a dependence of
$\phi$ on depth because then $\phi$ would have to decrease with depth,
which is unphysical.

There is no hysteresis in the
transition: measurements at different volume fractions can be made in
any order, without affecting the results.  However, the value of
$\phi$ at the transition depends weakly on the frictional
characteristics of the grains, which can change slowly with usage of
the beads \cite{schroeter:05}.

The penetration force is independent of the speed of the rod
(cf. Fig.~\ref{fig:transition}).  The absence of a dependence on
penetration speed is in agreement with the results of \cite{stone:04}
at $\phi$ = 0.59.  It also shows that the flow of water induced by the
penetrating rod does not alter the results, in agreement with
\cite{hill:05}.

\begin{figure}[t]
  \begin{center}
    \includegraphics[angle=-90,width=8.8cm]{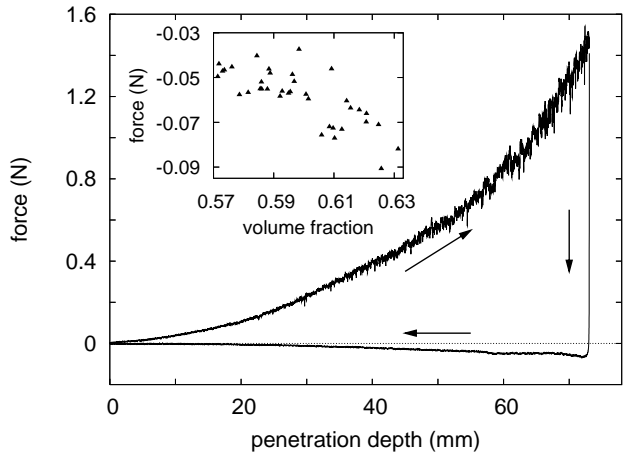}
    \caption{Forces measured during a full cycle of insertion
and withdrawal at a volume fraction $\phi$ = 0.602. 
The speed of the rod is 10 mm/min in both directions. 
The inset shows
the withrawal force measured at a depth of 70 mm for the same
experiments as in Fig.~\ref{fig:transition}.}
    \label{fig:withdrawal}
  \end{center}
\end{figure}

Typically, only about five percent of the force we measure is exerted on the
sides of the rod, as we see from Figure \ref{fig:withdrawal}, which measures the
force on withdrawal. The measurements during withdrawal also suggest the phase transition,
though it is not well defined since the forces are so much smaller.

\begin{figure}[t]
  \begin{center}
    \includegraphics[angle=-90,width=8.8cm]{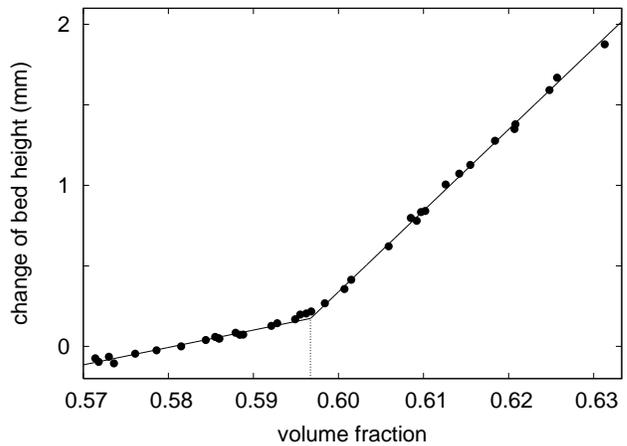}
    \caption{Change in the height of the bed relative to the original
      bed height for a rod 
      inserted to a depth of 60 mm. The transition occurs at $\phi =
      0.597$. The bed is slightly deformed near the rod but not
      sufficiently to change the average height of the bed.}
    \label{fig:bedheight}
  \end{center}
\end{figure}

The transition between distinct phases indicated by
Fig.~\ref{fig:transition} should be accompanied by changes in other
properties. Indeed, measurements of the average height of the bed as a
function of $\phi$ reveal, as Fig.~\ref{fig:bedheight} shows, a
transition at the same $\phi$ as in the force measurements.

{\it Discussion.---} A system similar to ours whose phase transitions
have been studied is a collection of colloidal particles prepared at
different $\phi$ by centrifugation \citep{pusey:86}. The colloidal
system has been characterized by a freezing density $\phi_f=0.407$ and
a melting density $\phi_m=0.442$, such that for $\phi < \phi_f$ the
system is a fluid, for $\phi > \phi_m$ the system is a crystalline
solid, and for $\phi_f < \phi < \phi_m$ the system is a mixture of the
two states. This phase behavior (but with $\phi_f=0.494$ and
$\phi_m=0.545$) was found earlier in Monte Carlo and molecular
dynamics simulations for frictionless hard spheres~\cite{frenkel:02}.

There are differences between a colloidal system and
our static granular system. For example, the colloidal system has
pure solid and fluid phases separated by a coexistence region with
a sharp transition at each end, while the granular system exhibits
only one transition in our experiment.  Another 
difference is that the transition in the colloidal system
corresponds to a change of symmetry from disordered to
crystalline, while the transition in the granular system is marked
by a change in the resistance to shear. However, a
recent consideration of the transition in the equilibrium hard
sphere model \citep{bowen:06} suggests that the symmetry change
may be hiding the fundamental mechanism in that model, a type of
geometric constraint, which may also underlie the phases of the
granular system.

The volume fraction at which we observe a transition coincides with earlier
observations of interesting behavior. 
In \cite{schroeter:05} volume
fluctuations around a static steady state were measured using a series
of identical flow pulses in a fluidized bed, and those fluctuations
were found to exhibit a parabolic minimum at volume fractions between
0.587 and 0.596, depending on the surface roughness of the
beads. Arguments based on the central limit theorem showed that this
minimum corresponds to a minimal number of beads being contained in a
statistically independent region, which is tantamount to a minimum in
the correlation length.  Another study found
that if a fluidized bed is slowly defluidized, it exhibits a behavior
similar to that of a supercooled liquid, with an arrest transition at
$\phi$ = 0.594 \cite{goldman:06}. At the arrest transition $\phi$ became nearly
independent of the flow rate and the correlation length went to zero.
A similar result has been obtained in an analysis of the Delauney tesselation
of tomograms of sphere packings: at volume fractions between 0.58 and
0.60 re-adjustments involving only a single sphere become impossible and
any dynamics requires collective and correlated motion of larger
sets of spheres \cite{aste:05}.  And finally, as indicated by Figure
\ref{fig:bedheight}, 
our transition takes place at the critical state
density of soil mechanics\cite{nedderman:92}; at densities below this
the granular sample collapses under stress, while at densities above this it expands
under stress.

In conclusion, we have taken advantage of a method to produce
beds of granules with well-defined volume
fractions   to search
for a phase transition as a function of volume fraction.  The
transition revealed by our measurements of penetration force and bed
height should help in the understanding of granular behavior, for instance, in 
avalanches \cite{evesque:91}. The existence of a well-defined
transition between two phases also 
suggests the appropriateness of a statistical approach 
to understanding the phases of granular matter.

We thank Persi Diaconis, Pierre Evesque, Daniel Goldman, W.D. McCormick, and Peter
Schiffer for helpful discussions.  Schr\"oter and Swinney were
supported in part by the Robert A. Welch Foundation; Radin was
supported in part by NFS grant DMS-0352999.

\end{document}